\documentclass[10pt,conference]{IEEEtran}
\IEEEoverridecommandlockouts
\usepackage{cite}
\usepackage{amsmath,amssymb,amsfonts}
\usepackage{algorithmic}
\usepackage{graphicx}
\usepackage{textcomp}
\usepackage{xcolor}
\usepackage{makecell}
\usepackage[linesnumbered,ruled,vlined]{algorithm2e}
\usepackage{caption}
\usepackage{color}
\usepackage{stfloats}
\usepackage{booktabs}
\usepackage{multirow}
\usepackage{verbatim}
\usepackage{bm}
\usepackage{url} 
\usepackage{framed} 
\usepackage{threeparttable}    
\usepackage{balance}   

\usepackage{amsmath}
\allowdisplaybreaks[4] 

\SetCommentSty{mycommfont}
\SetKwInput{KwInput}{Input}                
\SetKwInput{KwOutput}{Output}              

\newcommand{\model}{{\emph{MRAM}}\xspace}
\newcommand{\fine}{{\emph{FineLocator}}\xspace}
\newcommand{\blia}{{\emph{BLIA 1.5}}\xspace}
\newcommand{\bliz}{{\emph{Blizzard}}\xspace}

\newcommand{\graph}{{\text{code revision graphs}}\xspace}
\newcommand{\smnn}{{\text{SMNN}}\xspace}
\newcommand{\menn}{{\text{MENN}}\xspace}
\newcommand{\flnn}{{\text{FLNN}}\xspace}
\newcommand{\bff}{{\text{bug-fixing features}}\xspace}
\newcommand{\msf}{{\text{method structured features}}\xspace}

\newcommand{\bffs}{{\text{bug fixing frequency score}}\xspace}
\newcommand{\bfrs}{{\text{bug fixing recency score}}\xspace}

\newcommand{\cev}[1]{\reflectbox{\ensuremath{\vec{\reflectbox{\ensuremath{#1}}}}}}

\newcommand{\etal}{\hbox{\emph{et al.}}\xspace}
\newcommand{\eg}{\hbox{\emph{e.g.}}\xspace}
\newcommand{\ie}{\hbox{\emph{i.e.}}\xspace}

\def\BibTeX{{\rm B\kern-.05em{\sc i\kern-.025em b}\kern-.08em
	T\kern-.1667em\lower.7ex\hbox{E}\kern-.125emX}}
	
\begin{document}
\title{Locating Faulty Methods with a Mixed RNN and Attention Model}

\author{
	\IEEEauthorblockN{Shouliang Yang, Junming Cao, Hushuang Zeng, Beijun Shen\IEEEauthorrefmark{1}, Hao Zhong
	\thanks{\IEEEauthorrefmark{1} Beijun Shen is the corresponding author.}}
	\IEEEauthorblockA{
		\textit{School of Electronic Information and Electrical Engineering} \\
		\textit{Shanghai Jiao Tong University}\\
		Shanghai, China \\
		\{ysl0108, junmingcao, zenghushuang, bjshen, zhonghao\}@sjtu.edu.cn
}
}

\maketitle

\begin{abstract}
	\textit{IR-based fault localization} approaches achieves promising results when locating faulty files by comparing a bug report with source code. Unfortunately, they become less effective to locate faulty methods. We conduct a preliminary study to explore its challenges, and identify three problems: \textit{the semantic gap problem}, \textit{the representation sparseness problem}, and \textit{the single revision problem}.
	
	To tackle these problems, we propose \model, a mixed RNN and attention model, which combines \bff and \msf to explore both implicit and explicit relevance between methods and bug reports for method level fault localization task.
	The core ideas of our model are: 
	(1) constructing \graph from code, commits and past bug reports, which reveal the latent relations among methods to augment short methods and as well provide all revisions of code and past fixes to train more accurate models;
	(2) embedding three \msf  (token sequences, API invocation sequences, and comments) jointly with RNN and soft attention to represent source methods and obtain their implicit relevance with bug reports;
	and (3) integrating multi-revision \bff, which provide the explicit relevance between bug reports and methods, to improve the performance.
	
    We have implemented \model and conducted a controlled experiment on five open-source projects. Comparing with state-of-the-art approaches, our \model improves MRR values by 3.8-5.1\% (3.7-5.4\%) when the dataset contains (does not contain) localized bug reports. Our statistics test shows that our improvements are significant.

\end{abstract}


\section{Introduction}

With the increase of software’s complexity and scale, endless emerging bugs consume a huge amount of development time and effort.
It is rather difficult and time-consuming for developers to locate faulty code precisely, especially when a project has many source files. Program debugging and fault localization can take up to $50\%$ of developer's working hours and $25\%$ of total software development costs~\cite{britton2013reversible}. As a result, efficient fault localization techniques are desirable to reduce the costs.

When programmers encounter bugs, they often report them through issue trackers. Given a bug report, \textit{information retrieval (IR)-based} fault localization approach reduces the localization problem to a search problem: bug reports are queries, and code snippets are answers\cite{buglocator, fse2014, BLIA, FineLocator}. 
Most existing works~\cite{buglocator,fse2014,DNNLoc} localize faults at the granularity of file level. 
However, according to a survey on desirable fault localization approaches~\cite{kochhar2016practitioners}, $51.81\%$ programmers prefer to locate faulty methods, and only $26.42\%$ of programmers are satisfied with locating faulty files. 
Therefore, approaches \cite{BLIA, FineLocator} start to explore locating faulty methods, but their effectiveness is less impressive.

As the number of methods is much higher than the number of source files, intuitively, locating faulty methods is more challenging than locating faulty files. 
Besides this issue, we identify three challenging problems in locating faulty methods (Section~\ref{sec:pre}):
\textbf{(1) The semantic gap problem.}
Most IR-based approaches treat both bug reports and source files as natural language texts, and
use bag-of-words features to correlate bug reports and source files in the same lexical feature space.
However, the terms in bug reports are written in natural languages, and may differ from the tokens in source code. Thus, there is a semantic gap between bug reports and source methods.
Even though embedding techniques~\cite{FineLocator} have been used to encode bug reports and source methods into the same space, they usually ignore the syntax and semantics of code, and thus still suffer from this problem.
\textbf{(2) The representation sparseness problem.} Short methods are popular in software projects.
For example, in Tomcat, we find that more than 70\% of methods have fewer than 5 statements, and more than 45\% of methods have fewer than 3 statements.
With only several statements, short methods are ambiguous, and difficult to be matched against a bug report. \textbf{(3) The single revision problem.}
A source method in a software repository often has a long revision history, but the prior approaches (\eg \cite{BLIA},\cite{Blizzard}) typically analyze only the latest code to locate faulty methods. This problem results in lacking of sufficient historical data to train an accurate model.

To resolve the three problems, we propose a mixed neural network called \model for locating faulty methods.
\model combines \msf and \bff through \graph, and it identifies the faulty methods of a bug report accurately with 
recurrent neural network (RNN)~\cite{kombrink2011recurrent}, 
soft attention mechanism \cite{cho2014properties}, 
and multi-layer perception (MLP)\cite{gardner1998artificial}.
Compared with the state-of-the-art approaches \cite{FineLocator, BLIA, Blizzard}, 
our experimental results show that \model significantly improves their effectiveness. 
For example, the average MAP is increased from 
0.0305 of Youm \etal~\cite{BLIA}, 0.0249 of Rahman \etal~\cite{Blizzard}, and 0.0311 of Zhang \etal~\cite{FineLocator} to our 0.0606. 
We have used the Mann-Whitney U Test\cite{mann1947test} to compare
our top recommendations with those of the prior approaches, and the results show that our improvements are significant. \model is able to make these improvements,
because it considers both implicit and explicit relevance between bug reports and methods, 
and is the first to enrich short methods with their past fixes.

This paper makes the following contributions:
\begin{enumerate}
    \item \textbf{A mixed RNN and Attention model}.
    \model uses RNN and soft attention to merge extra structural information of source methods to obtain their implicit relevance to bug reports, which solves the semantic gap problem.
	Then, the implicit relevance along with three \bff, which provide the explicit relevance, 
	are combined to calculate the correlations between bug reports and methods.
    
    \item \textbf{Code revision graphs}. 
    In this paper, we build code revision graphs from past code commits and bug reports. As our graphs reveal latent relations among methods, they are useful to expand short methods, and thus resolve the representation sparseness problem.
    Furthermore, we extract accurate \bff from our graphs. As our features are extracted from the revision history of a method than its latest version, we resolve the single revision problem.
   
    \item \textbf{An open source tool and evaluations}. We have released our tool and dataset on Github\footnote{https://github.com/OrthrusFL/MRAM} and evaluated it on five large scale open source Java projects.
    The results show that it could localize bugs at the granularity of methods more precisely than baselines, 
    achieving the MAP of 0.0606 and the MRR of 0.0764. 
    We perform a series of empirical experiments to show the improvement of \model over the prior approaches.

\end{enumerate}

\section{Background}



\subsection{Bidirectional RNN for Sequence Embedding}
Embedding is a widely used technique to boost the performance of natural language processing (NLP) tasks. 
It learns to map entities like words, phrases, sentences, or images to vectors of real numbers in a dense vector space.
Word embedding is a typical embedding technique that produces a fixed-length vector for a unique word, where similar words are close to each other in the vector space.
It can be represented as a function $E: \mathcal{V} \rightarrow \mathbb{R}^{d}$ that takes a word $w$ in vocabulary $\mathcal{V}$ and maps it to a vector in a $d$-dimensional vector space. 

Convolutional neural network (CNN) and RNN are the two main types of deep neural network architectures for embedding technique. 
Compared to CNN, RNN performs better at modeling unit sequence.
Trained to predict the next symbol in a sequence, RNN models the probability distribution of a sequence.
Therefore, RNN achieves impressive results in sequence embedding.
Among RNN variants, bidirectional RNN (BRNN)~\cite{schuster1997bidirectional} consists of two unidirectional RNNs in opposite directions, therefore can utilize information from past and future states simultaneously to capture the context of the sequence.

In a sentence $T=t_{1},...,t_{N_{T}}$,
each word $t_{i}$ is mapped to a vector $\bm{t_{i}}$ with $d$-dimensional by word embedding~\cite{mikolov2013distributed}:
    \begin{equation}
        \bm{t_{i}} = Embedding(t_{i}).
    \end{equation}

Then each of the vector $\bm{t_{i}}$ along with the preceding hidden state $\vec{h}_{i-1}$
updates the hidden state of forward RNN from front to back; 
and then updates the hidden state of backward RNN from back to front
with the following hidden state $\cev{h}_{i+1}$. 
Next, $\vec{h_{i}}$ and $\cev{h_{i}}$ are concatenated to present current hidden state:

\begin{equation}
    \begin{split}
    	&\vec{h}_{i} = \tanh(\bm{W_{M}^{f}}[\vec{h}_{i-1}; \bm{t_{i}}]) , \\
		&\cev{h}_{i} = \tanh\bm{(W_{M}^{b}}[\cev{h}_{i+1}; \bm{t_{i}}]) ,  \\ 
		&\bm{O_{i}} = \bm{W_{M}} [\vec{h}_{i};\cev{h}_{i}] + b , \\
	\end{split}
\end{equation}
where $[a;b] \in \mathbb{R}^{2d} $ represents the concatenation of two vectors,
$\bm{W_{M}^{f}} \in \mathbb{R}^{2d \times d} $, $\bm{W_{M}^{b}} \in \mathbb{R}^{2d \times d} $,
$\bm{W_{M}} \in \mathbb{R}^{2d \times d} $ and b are the matrices of trainable parameters,
$\vec{h}_{i} \in \mathbb{R}^{d} $ and $\cev{h}_{i} \in \mathbb{R}^{d}$ represent the forward and backward hidden states for timestep $i$,
$\bm{O_{i}} \in \mathbb{R}^{d} $ is the output of BRNN for token $t_{i}$,
and $\tanh$ is an activation function of the BRNN.

Finally, all the hidden states are fed into an output layer to get the vector representation of the sentence $T$ for a specific task.
A typical output layer is a maxpooling function\cite{kim2014convolutional}:

\begin{equation}
   	\bm{m} = \mbox{maxpooling}([\bm{O_{1}},...,\bm{O_{N_{M}}}]) .\\
\end{equation}

\subsection{Soft Attention Mechanism for Text Expansion}

\begin{figure}[t]
    \centering{\includegraphics[scale=0.45]{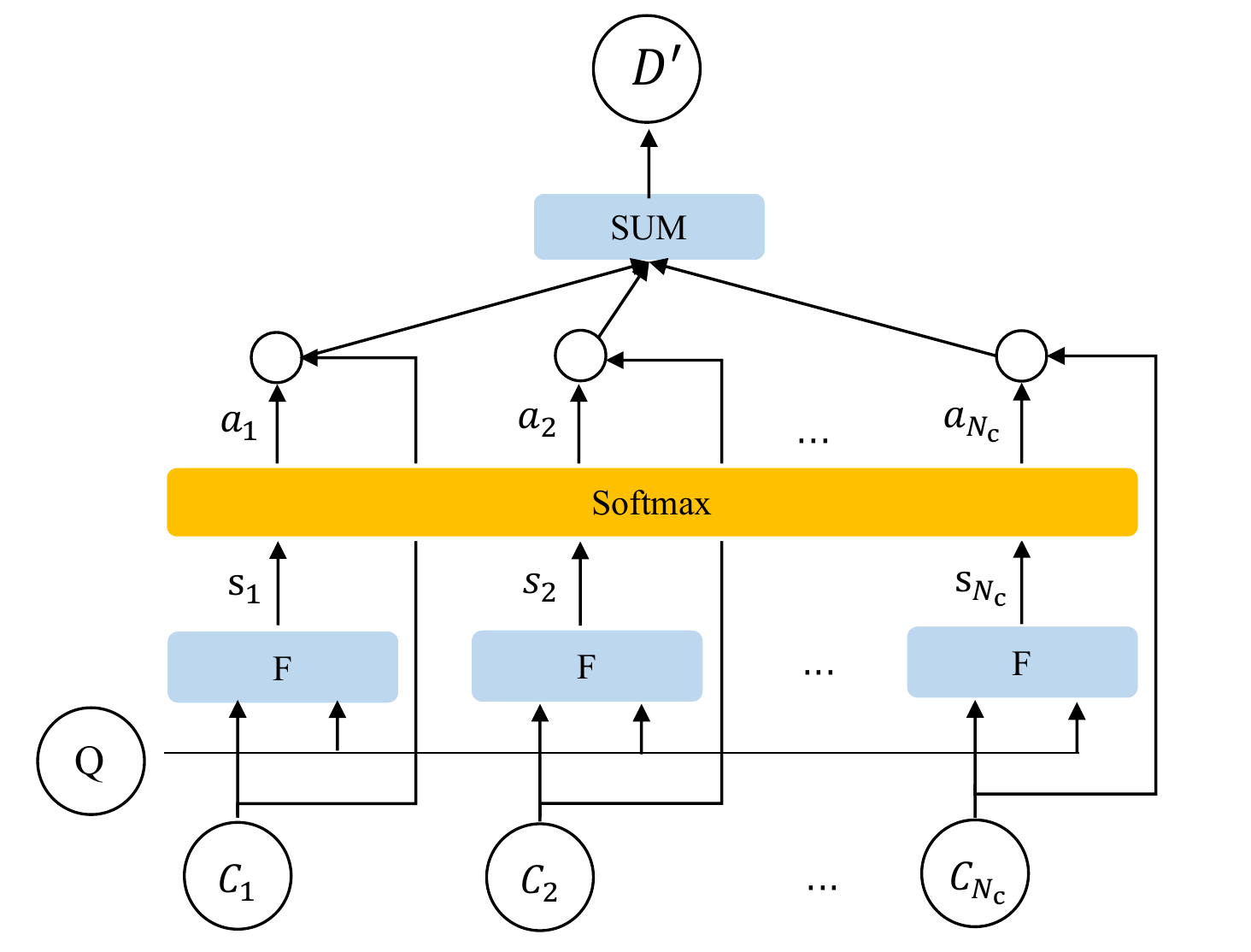}}\vspace{-3mm}
    \caption{A soft attention mechanism for text expansion}
    \label{fig_attention}
    \vspace{-4mm}
\end{figure}

Attention mechanism has become an important component in deep neural networks for diverse application domains.
The intuition of attention mechanism comes from the human biological systems: the visual processing system of humans tends to focus selectively on parts of the image while ignoring other irrelevant information \cite{luong2015effective}.
Among various attention mechanism, soft attention captures the global latent information of a sequence by using a weighted average of all hidden states of the input sequence. This approach is widely used in NLP tasks such as machine translation, text classification, and text expansion.

Take the text expansion as an example. Given a short text $Q$ and a collection of texts $C=C_{1}, C_{2},...,C_{N_{C}}$, where each text $C_{i}$ is relevant to $Q$.
Text expansion aims to expand $Q$ into a richer representation according to $C$.
Fig. \ref{fig_attention} shows an example of text expansion using soft attention.
The short text $Q$ and each text in $C$ are first mapped into two $d$-dimensional vectors by sequence embedding, and then are matched to get an attention score $s_{i}$:
\begin{equation}
    s_{i} = \mbox{tanh}(\bm{W_{q}}Q+\bm{W_{c}}C_{i}),
\end{equation}
where $\bm{W_{q}}$ and $\bm{W_{c}}$ are trainable parameters in soft attention mechanism.
Next, $s_{i}$ is passed to a softmax function for normalization to compute the attention weight $a_{i}$:
\begin{equation}
    a_{i} = \frac{e^{s_{i}}}{\sum_{j=1}^{N_{c}}e^{s_{j}}}.
\end{equation}
With softmax, the information retrieved from $C$ is computed as a weighted average for $C_{1}, C_{2},...,C_{N_{C}}$:
\begin{equation}
    D^{'} = \sum_{i=1}^{N_{C}}(a_{i}C_{i}).
\end{equation}
Finally, the retrieved information $D^{'}$ is used to augment the representation of $Q$.
A typical way is to concatenate the vector of $Q$ and $D^{'}$ as a new $Q$.

\begin{table*}[ht]
	\caption{Our dataset}\vspace{-2mm}
	\begin{center}
		\footnotesize
		\begin{tabular}{|c|c|c|ccc|cc|ccc|ccc|}
			\hline
			\multirow{2}*{Project}        
			& \multirow{2}*{Bug report}  
			& \multirow{2}*{Traceable report } %
			& \multicolumn{3}{c|}{Bug report type} 
			& \multicolumn{2}{c|}{Fixed method} 
			& \multicolumn{3}{c|}{Method}  
			& \multicolumn{3}{c|}{Short method} \\ 
			\cline{4-14}
			&  & &fully &partially &not &all &short &max &med &min &max &avg &min \\
			\hline
			AspectJ
			&530 &324 &157 &69 &304 &5.296 &1.868 &32,816 &19,713 &7,607 &20,613 &14,370 &5,913  \\
			\hline
			Birt
			&4,009 &2,792 &301 &224 &3,484 &7.075 &4.797 &100,625 &60,432 &14,441 &73,615 &22,080 &11,145    \\
			\hline
			JDT
			&6,058 &3,278 &1,101 &401 &4,556 &5.023 &1.668 &49,152 &24,714 &7,916 &34,516 &17,969 &6,244   \\
			\hline
			SWT
			&3,936 &928 &1,876 &472 &1,588 &5.320 &0.712 &16,165 &12,196 &6,088 &11,653 &8,017 &4,532   \\
			\hline
			Tomcat
			&979 &607 &9 &10 &960 &5.174 &2.040 &32,812 &19,314 &11,531 &23,162 &15,135 &8,692  \\
			\hline
		\end{tabular}		
		\label{dataset}
		\begin{tablenotes}
			\item[b] traceable report: the number of bug reports whose faulty methods appear in the latest code revision; 
			fixed method: the average of fixed methods per bug report; 
			method: the number of methods per code revision;
			short method: the number of methods with less than 5 sentences per code revision.
		\end{tablenotes}
	\end{center}
	\vspace{-7mm}
\end{table*}
\section{Preliminary Study} 
\label{sec:pre}
In this section, we conduct a preliminary study to analyze the problems of locating faulty methods.

\subsection{Setting}
\label{sec:pre:dataset}

The benchmark dataset created by Ye \etal~\cite{fse2014} from open-source Java projects
is used for our experiments.
This dataset is widely used in evaluating approaches that locate faulty files, and we make it adapt to the faulty methods locating task. 
On this dataset, we conduct a preliminary analysis of three problems: 
the semantic gap problem, the representation sparseness problem, and the single revision problem. We notice that the discussions of some bug reports already identify faulty files or methods~\cite{bias2014, bias2018}, and we call such bug reports \textit{localized bug reports}. As Kochhar \etal ~\cite{bias2014} did, we categorize the bug reports in our dataset into three categories:

\textit{1) fully localized}: faulty methods are identified;

\textit{2) partially localized}: faulty methods are partially identified;

\textit{3) not localized}: faulty methods are not identified.

Column ``Bug report type'' of Table~\ref{dataset} lists their numbers.

\subsection{Result}

An analysis of our dataset have revealed the main problems of
faulty methods locating task:

\textit{1) The semantic gap problem.} 
The terms used in bug reports written in natural languages evidently differ from the tokens used in methods written in programming languages. 
We adopt TF-IDF to measure their textual similarity.
The result shows that the average textual similarity between bug reports and their \textit{fixed} methods is 0.0153, while that between bug reports and their \textit{irrelevant} methods is 0.0149. 
Obviously, there is a big semantic gap between bug reports and methods, 
and a typical IR-based approach is not able to identify those faulty methods by matching textual similarity. 

\textit{2) The single revision problem.} 
As shown in column 2-3 of Table~\ref{dataset}, a lot of bug reports can not be localized when only using the latest code revision.
For example, there are only 928 bug reports whose faulty methods appear in the latest code revision in SWT project. 
One potential reason is that code changes occur more frequently in method level, and a great number of methods are deleted or changed across neighboring code revisions.

\textit{3) The representation sparseness problem.} 
As shown in column 7-8 of Table~\ref{dataset}, nearly 2/3 of all methods are short. It is difficult to handle the sparse representation when localizing short methods. Moreover, these short methods cannot be simply filtered out for their relatively large proportion of faulty methods.

\section{Code Revision Graph} \label{graph_all}

To learn more faulty localization knowledge from project historical data, we construct \graph from multi-revision code, commits, and past bug reports.

\begin{figure}[t]
    \centering{\includegraphics[scale=0.65]{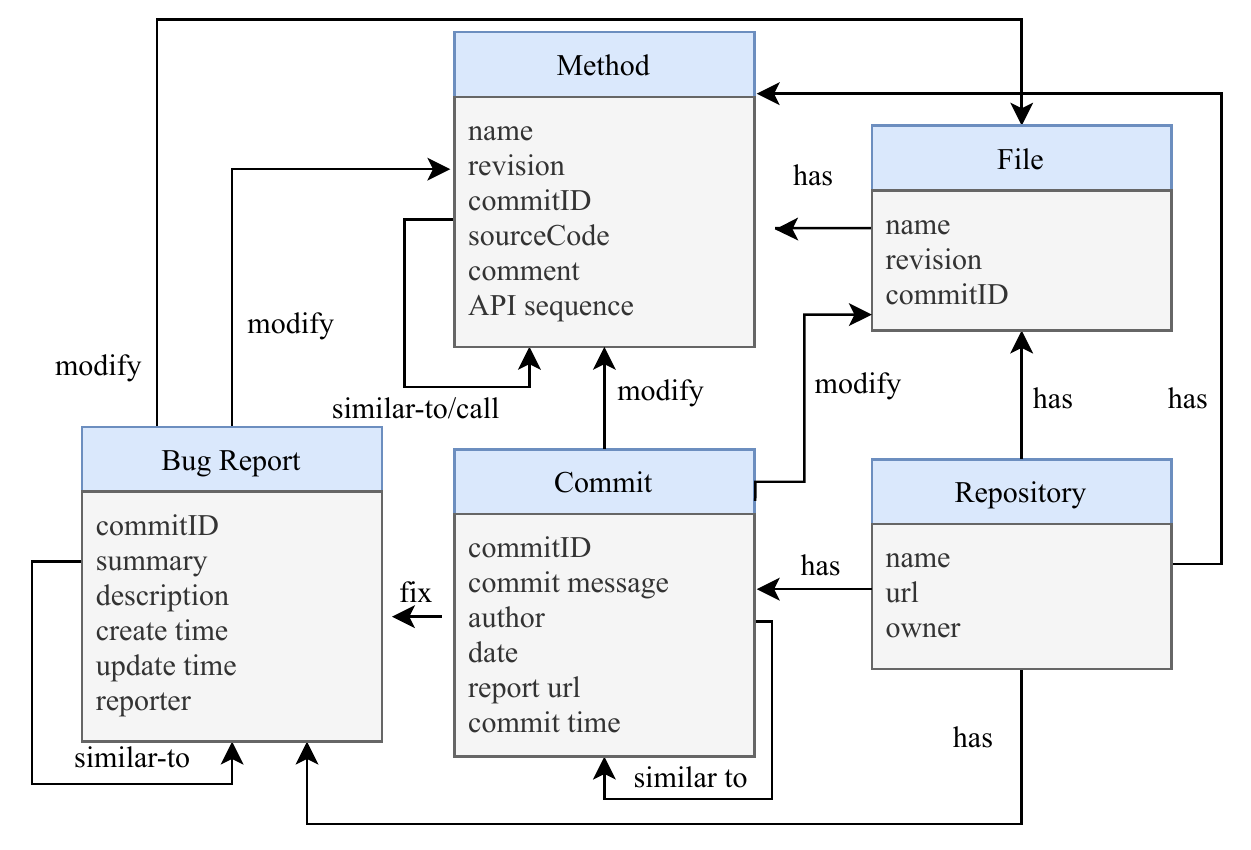}}\vspace{-2mm}
    \caption{The schema of our \graph}
    \label{ontology}
    \vspace{-4mm}
\end{figure}

As shown in Fig. \ref{ontology}, \graph define the following entities and relations between them:
repository, bug report, commit, file, and method.
The code entities (\ie file and method) have a ``revision" attribute.
The \graph offer the following benefits:

1) efficiently providing all revisions (not just the latest revision) of code,
bug reports, and commits to train the model;

2) alleviating the representation sparseness problem 
by deducting related methods to expand short methods;

3) calculating multi-revision \bff to improve the performance of fault localization.

\subsection{Constructing Graph} \label{construction}

Given a repository, we use Spoon~\cite{pawlak2016spoon},
an open-source library to parse Java source code to get code entities: files and methods,
and extract \textit{has}, \textit{modify} and \textit{call} relations.
The approach presented by Dallmeier \etal \cite{2007extraction} is used to obtain the \textit{fix} relations between code commits and bug reports.
In particular, we apply a structural-context similarity measurement, \textit{SimRank}\cite{simrank}, to identify the \textit{similar-to} relations.
Let $S^{b}_{ij}$ denotes the similarity between bug reports $b_{i}$ and $b_{j}$, $S^{m}_{ij}$ the similarity between methods $m_{i}$ and $m_{j}$, $M(b_{i})$ the set of methods modified by $b_i$, and $B(m_i)$ the set of bug reports that modified $m_i$.
\textit{SimRank} has the following three steps:
In step 1, $S^{b}_{ii}$ and $S^{m}_{ii}$ are set as 1 and will not be updated later. All other similarity scores are initialized as 0.
In step 2, similarity scores are updated iteratively (5 rounds of iteration) by the following equations:
\begin{equation}
	S^{b}_{ij}= \frac{C}{|M(b_{i})||M(b_{j})|}\sum_{k=1}^{|M(b_{i})|}\sum_{l=1}^{|M(b_{j})|}S^{m}_{kl}\;,
\end{equation}
\begin{equation}
	S^{m}_{ij}= \frac{C}{|B(m_{i})||B(m_{j})|}\sum_{k=1}^{|B(m_{i})|}\sum_{l=1}^{|B(m_{j})|}S^{b}_{kl}\;,
\end{equation}
where $C$ is the rate of decay as similarity flows in the \graph, and is set to 0.8.
In step 3, we pick the pairs that have a similarity score larger than 0.001 (except $S^{b}_{ii}$ and $S^{m}_{ii}$) as similar bug reports and methods.
Compared to content-based similarity measurements, \textit{SimRank} works better for short methods. Here, the \textit{similar-to} relations are computed directly from code revision graphs. As these relations define the similarity of the bug fixing history, they are useful to locate faulty methods.


It is expensive to build our graphs from scratch every time a new version is submitted. Instead, we update our graphs incrementally.
In particular, the modifications of a revision are classified into three types: \textit{additions}, \textit{deletions}, and \textit{modifications}.
The code entity of an \textit{addition} is added into \graph; that of an \textit{deletion} is marked as deleted; and that of a \textit{modification} is linked to its new revision with an \textit{update} relation.

\begin{figure}[t]
    \centering{\includegraphics[scale=0.5]{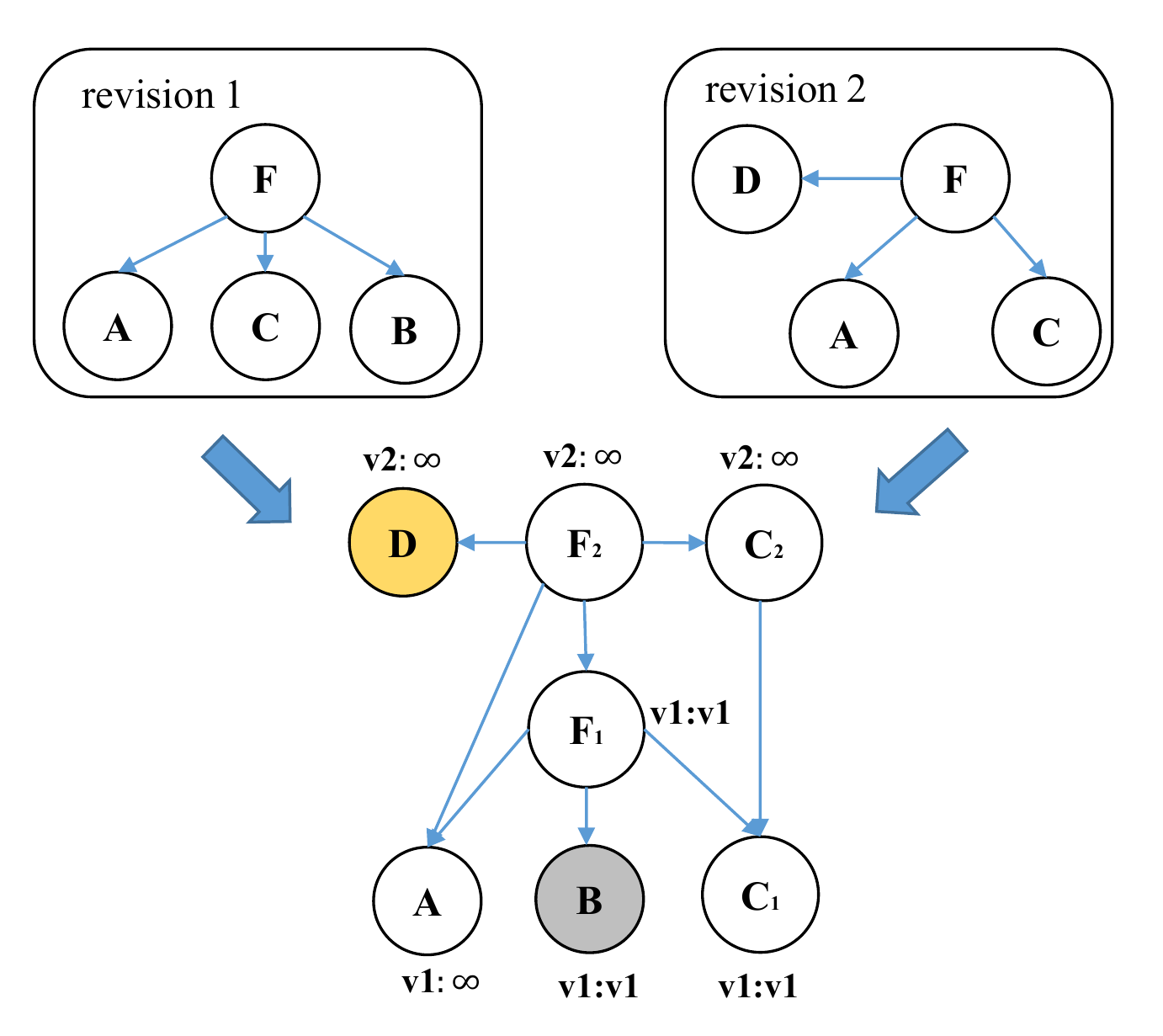}}\vspace{-2mm}
	\caption{An example of merging code entities from two neighboring revisions.
	The yellow circle represents an added entity and the gray circle represents a deleted entity.}
    \label{graph}
    \vspace{-2mm}
\end{figure}

Fig. \ref{graph} illustrates how we merge code entities from two neighboring revisions.
In revision 1, file F contains three methods: A, B, C; in revision 2, method B is deleted from F, method D is added into F while
method C is modified. In the code revision graph, a method before and after modification are treated as two different entities and
identified by the ``revision'' attribute. This solution is limited to a linear history of revisions.
In this way, it takes only several minutes to maintain and update \graph.

\subsection{Calculating Bug-Fixing Feature} \label{feacalculate}
Code change history is an important metric for detecting defects \cite{rahman2013and}.
Therefore, we select three types of \bff which are widely used in fault prediction models to improve the performance of fault localization:
collaborative filtering score, \bffs, and \bfrs.
Further, we revise the collaborative filtering score to fit the faulty methods locating task,
since it 
heavily depends on the \textit{fix} relations
which may not be sufficient in a software repository.
The algorithms to calculate the \bff are introduced as follows.

\subsubsection{Revised Collaborative Filtering Score (\textit{rcfs})}
\textit{rcfs} calculates the relevant values between
a new given bug report $b_{i}$ and all source methods $M$ in its before-fix code revision according to the history of previous code revision \cite{buglocator}.
Its calculation relies on \textit{similar-to} relations between ${b_i}$ and its previous bug reports $B_{prev}(b_{i})$, and \textit{fix} relations between $B_{prev}(b_{i})$ and $M$.
However, there are no existing \textit{similar-to} relations of ${b_i}$ because \textit{SimRank} could only be used for localized bug reports.
Moreover, our \graph do not have many \textit{fix} relations, and may underestimate the similarity values.
Therefore, we propose Algorithm \ref{alg1} to compute \textit{rcfs} for each $m$ in $M$ when given a new bug report $b_{n}$.
In Step 1, we revise the cosine similarity values between $b_n$ and $b{_i} \in B_{prev}$ with $S^{b}$.
In Step 3, \textit{cfs} is revised into \textit{rcfs} via $S^{m}$.

\begin{algorithm}
	\DontPrintSemicolon
	\footnotesize
	  \KwInput{$b_{n}$, $M$, $S^{b}_{ij}$, $S^{m}_{ij}$, $B_{prev}(b_{i})$}
	  \KwOutput{ $rcfs_{m_{i}}$, for ${m}_{i} \in M$}
	  \tcc{Step 1: get $S^{b}_{ni}$, for $b_i \in B_{prev}(b_{n})$}
	  \For{$b_{i} \in B_{prev}(b_{n})$}{
		  $S^{b}_{ni}$ = $\sum_{j=1}^{|B_{prev}(b_{n})|}S^{b}_{ij}*cosSim(b_{j},b_{n}) + cosSim(b_{i}, b_{n})$ \;
	  }
	  \tcc{Step 2: get $cfs_{m_{i}}$, for $m_{i} \in M$}
	  $M_{b_{i}}$ denotes methods that are modified by $b_i$ \;
	  \For{$m_{i} \in M$}{
		\For{$b_{j} \in B^{prev}$}{
			\If{$m_{i}  \in M_{b_{j}}$}{
				add $b_{j}$ into $B_{m_{i}}$
			}
		  }
		$cfs_{m_{i}}$= $\sum_{j=1}^{|B_{m_{i}}|}{ S^{b}_{nj} * \frac{1}{|M_{b_{j}}|}, b_{j} \in B_{m_{i}}} $ \;
	  }

	  \tcc{Step 3: revise $cfs_{m_{i}}$ with $S^{m}$}
	  \For{$m_{i} \in M$}{
		$rcfs_{m_{i}}$= $\sum_{j=1}^{|M|} cfs_{m_{j}}*S^{m}_{ij} + cfs_{m_{i}}$\;
	  }
	\caption{The procedure of calculating \textit{rcfs}}
	\label{alg1}
	\vspace{-1mm}
\end{algorithm}

\subsubsection{Bug Fixing Recency Score (\textit{bfrs})}
We calculate \textit{bfrs} based on the assumption that changing code may introduce more faults, which has been proven correct at locating faulty files \cite{kim2007predicting}.
That is to say, a method that fixed recently has a high probability to be a faulty method in the near future.
Given a bug report $b_{i}$ and a method $m_{i}$, $b_{i}^{p}$ is the bug report that is most recently fixed in $m_{i}$, then \textit{bfrs} of $m_{i}$ for $b_{i}$ can be calculated as follows.
\begin{equation}
	bfrs(m_{i})=\frac{1}{k+1}\;,
\end{equation}
where $k$ counts the number of the month between $b_{i}$ and $b_{i}^{p}$.
If $m_{i}$ has never been fixed before $b_{i}$ is reported, then \textit{bfrs} of $m_{i}$ is $0$.

\subsubsection{Bug Fixing Frequency Score (\textit{bffs})}
\textit{bffs} measures how often a method is fixed.
It is based on the assumption that the more frequently a method has been fixed, the more likely it is the method causing the fault.
Given a bug report $b_{i}$ and a source method $m_{i}$, the \textit{bffs} of $m_{i}$ is scored as the number of times that $m_{i}$ has been fixed before $b_{i}$ is reported.

\section{\model}

We propose a  neural network named \model for locating faulty methods.
This section describes its network architecture, detailed design of main components, and model training.

\subsection{Overall Architecture}
\begin{figure*}[ht]
    \centering{\includegraphics[scale=0.53]{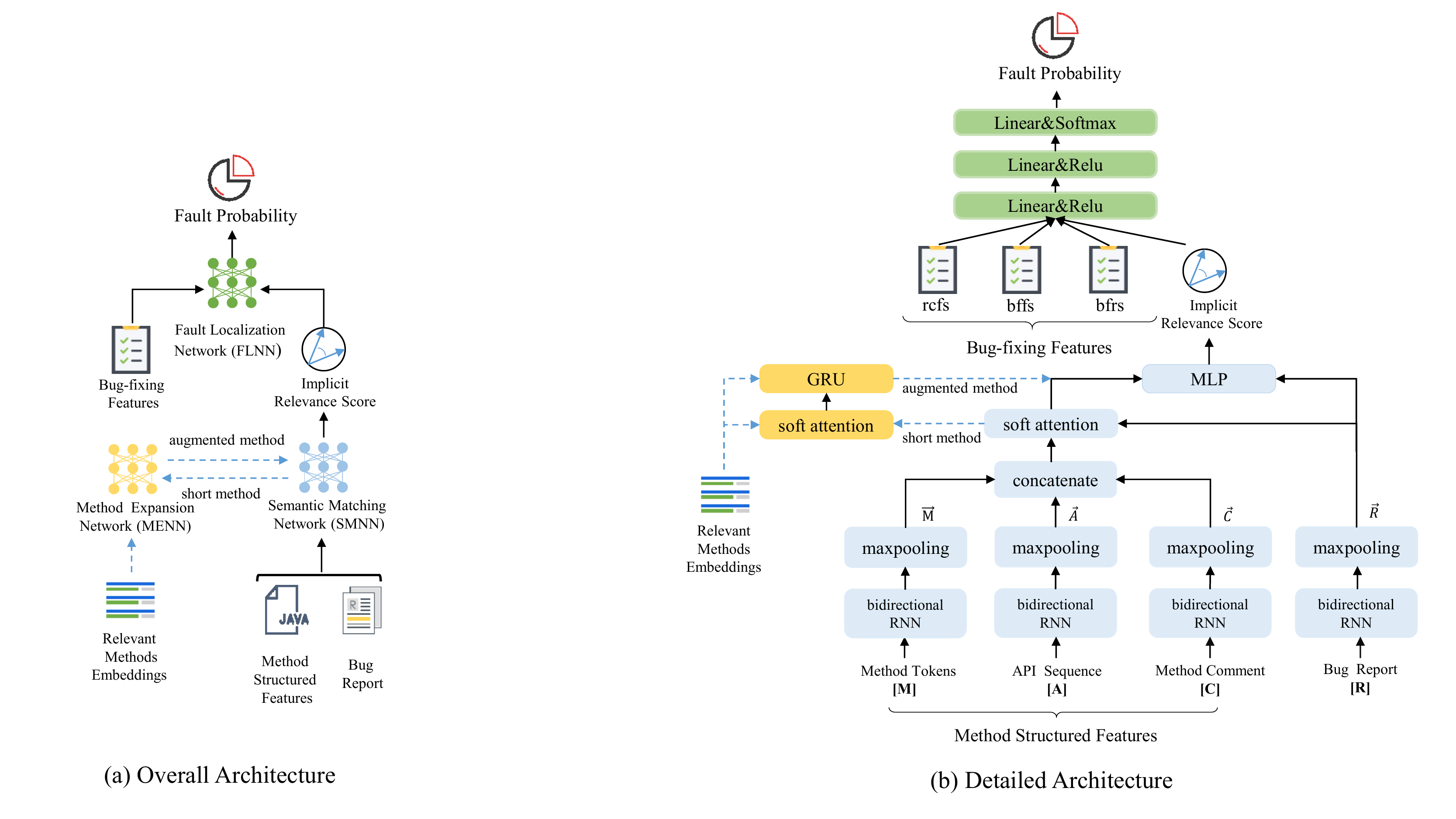}}\vspace{-1mm}
    \caption{The structure of our mixed RNN and attention model}
    \label{example}
    \label{fig_model}
    \vspace{-3mm}
	\end{figure*}

Fig. \ref{fig_model} shows the overall architecture of \model,
consisting of three main components:

1) \smnn (semantic matching network), which captures both semantic and structural information of a source method with bidirectional RNNs and soft attention. In this way, the source method can be matched with the bug report accurately in a unified vector space.

2) \menn (method expansion network), 
which enriches the representation of a method with short length by retrieving the information from its relevant methods;

3) \flnn (fault localization network), which predicts the fault probability of a method by combining both its 
implicit reference and explicit relevance to the bug report.

\subsection{Semantic Matching Network}\label{smnn}

To bridge the gap between natural language and programming language, semantic matching network (\smnn) embeds bug reports and source methods into a unified vector space. Different from bug reports, methods provide additional structural information besides the lexical terms and shouldn't be treated as plain texts. Therefore \smnn represents the source method by its three \msf:
the token sequence, the API invocation sequence, and the method comment.
Inspired by the attention mechanism for text expansion, we use it to embed multi-sources of information jointly.
Specifically, the three features along with the bug report are first embedded as a vector representation by bidirectional RNN respectively;
then, taking the bug report as a reference,
a soft attention mechanism is used to retrieve crucial information from vectors of three features to represent the whole method.
Thereafter the vector of the bug report is matched against the vector of the method by an MLP.

Consider a method $\mathcal{S}$=$[M, A, C]$ and a bug report $\mathcal{R}$,
where $M$=$m_{1},...,m_{N_{M}}$ is the source method represented as a sequence of $N_{M}$ split tokens;
$A$=$a_{1}, ...,a_{N_{A}}$ is the API sequence with $N_{A}$ consecutive API method invocations;
$C$=${c_{1}, ...,c_{N_{C}}}$ is the method comment represented as a sequence of $N_{C}$ split tokens;
and $\mathcal{R}$=${r_{1}, ...,r_{N_{R}}}$ is the bug report represented as a sequence of $N_{R}$ split tokens.
They are first embedded into a $d$-dimensional vector individually by
a bidirectional RNN\cite{gru} with a fully connected layer and a maxpooling layer:

\begin{equation}
	\begin{split}
		&\vec{h}_{t} = \tanh(\bm{W_{M}^{f}}[\vec{h}_{t-1}; \bm{m_{t}}]) , \\
		&\cev{h}_{t} = \tanh\bm{(W_{M}^{b}}[\cev{h}_{t+1}; \bm{m_{t}}]) ,  \\ 
		&\bm{O_{t}} = \bm{W_{M}} [\vec{h}_{t};\cev{h}_{t}] + b , \\
		&\bm{m} = \mbox{maxpooling}([\bm{O_{1}},...,\bm{O_{N_{M}}}]) ,\\
	\end{split}
\end{equation}
where $\bm{m_{t}} \in \mathbb{R}^{d}$ is the embedding vector of token $m_{t}$,
$[a;b] \in \mathbb{R}^{2d} $ is the concatenation of two vectors,
$\vec{h}_{t} \in \mathbb{R}^{d} $ and $\cev{h}_{t} \in \mathbb{R}^{d}$ are the forward and backward hidden states for timestamp $t$,
$\bm{O_{t}} \in \mathbb{R}^{d} $ is the output of bidirectional RNN with a fully connected layer for token $m_{t}$,
$\bm{W_{M}^{f}} \in \mathbb{R}^{2d \times d} $, $\bm{W_{M}^{b}} \in \mathbb{R}^{2d \times d} $ and
$\bm{W_{M}} \in \mathbb{R}^{2d \times d} $ are the matrices of trainable parameters,
$tanh$ is an activation function of the bidirectional RNN, 
and $b$ is the bias for the fully connected layer.
The token sequence $M$ is embedded as a $d$-dimensional vector $\bm{m}$.
Likewise, the API sequence $A$ is mapped to a vector $\bm{a}$ in the same vector space,
the method comment $C$ is mapped to a vector $\bm{c}$,
and the bug report $R$ is also mapped into a vector $\bm{r}$.\

Then, with reference to the vector of bug report $\bm{r}$, a soft attention is used to dynamically select the key information
from $\bm{m}$, $\bm{a}$, and $\bm{c}$ to represent
the source method:
\begin{equation}
	\begin{split}
		& s_{i} = \mbox{tanh}(\bm{W_{v}}\bm{v_{i}} + \bm{W_{r}}\bm{r}),\forall \bm{v_{i}} \in \{\bm{m},\bm{a},\bm{c}\} , \\
		& a_{i} = \frac{e^{s_{i}}}{\sum_{j=1}^{3}e^{s_{j}}} ,\\
		& \bm{s} = a_{1}\bm{v_{1}}+a_{2}\bm{v_{2}}+a_{3}\bm{v_{3}} ,
	\end{split}
\end{equation}
where $\bm{W_{v}}$ and $\bm{W_{r}}$ are the matrixes of trainable parameters in soft attention.
The output vector $\bm{s}$ represents the final embedding of the source method.
Finally, the vector of source method $\bm{s}$ will be matched against the vector of bug report $\bm{r}$ by an MLP:
\begin{equation}
	e = \bm{W_{2}} \sigma(\bm{W_{1}} \left[\bm{s};\bm{r} \right]+b_{1})+b_{2},
\end{equation}
where $W_{i}$ and $b_{i}$ are the weight and bias for $i^{th}$ MLP layer.
$e$ is the explicit matching score between $\bm{s}$ and $\bm{r}$.
The higher $e$ is, the more closely $\bm{s}$ is related to $\bm{r}$.

\subsection{Method Expansion Network}\label{menn}
A common practice to expand a short method is to enrich it with external information, 
which can be harvested from its relevant methods.
Therefore if the method $\mathcal{S}$ is short of length,
\menn will enrich its representation by integrating the retrieved information from
its relevant methods $M_{rel}(\mathcal{S})$.
$M_{rel}(\mathcal{S})$ includes two types of methods:
$M_{sim}(\mathcal{S})$ -- a set of methods with high SimRank similarity scores for $\mathcal{S}$,
and $M_{call}(\mathcal{S})$  --  a set of methods that $\mathcal{S}$ calls.
Both $M_{sim}(\mathcal{S})$ and $M_{call}(\mathcal{S})$ are collected from \graph.
Each $N_{i} \in M_{rel}(\mathcal{S})$ is also embedded into a vector just like $S$.

\menn first uses a soft attention mechanism to retrieve the relevant information for $S$ from $M_{rel}(\mathcal{S})$ automatically:
\begin{equation}
	\begin{split}
		& s_{i} = \mbox{tanh}(\bm{W_{n}}\bm{n_{i}} + \bm{W_{s}}\bm{s}) , \\
		& a_{i} = \frac{e^{s_{i}}}{\sum_{j=1}^{|M_{rel(S)}|}e^{s_{j}}} , \\
		& \bm{u} =  \sum_{i=1}^{|M_{rel(m)}|} {a_{i} \bm{n_{i}}} ,
	\end{split}
\end{equation}
where $\bm{s}$ is the vector of $\mathcal{S}$, $\bm{n_{i}}$ is the vector of $N_{i}$,
$\bm{W_{n}}$ and $\bm{W_{s}}$ are the matrices of trainable parameters in soft attention,
$s_{i}$ is the attention score for $n_{i}$, $a_{i}$ denotes the attention weight over $n_{i}$, and $\bm{u}$ is the relevant information retrieved from $M_{rel}(\mathcal{S})$ for $\mathcal{S}$.

Then, instead of directly adding $\bm{s}$ and $\bm{u}$ \cite{FineLocator}, which may bring about noisy data,
a gated recurrent unit (GRU) is employed to integrate the retrieved information $\bm{u}$ into $\bm{s}$.
As a result, the embedding of expanded method $\bm{\hat{s}}$ becomes richer and denser than the original embedding $\bm{s}$.
\begin{equation}
	\begin{split}
	&\bm{q} = \sigma (\bm{W_{q}}\bm{s} + \bm{U_{q}}\bm{u}) ,\\
	&\bm{r} = \sigma (\bm{W_{r}}\bm{s} + \bm{U_{r}}\bm{u}) ,\\
	&\bm{\hat{u}} = \tanh (\bm{W_{\hat{u}}} \bm{s} + \bm{r} \circ \bm{U_{\hat{u}}} \bm{u}) ,\\
	&\bm{\hat{s}} = \left( 1-\bm{q} \right) \circ \bm{s} + \bm{q} \circ \bm{\hat{u}} ,
	\end{split}
\end{equation}
where $\circ$ denotes element-wise multiplication, $\sigma$ is the logistic sigmoid activation function,
$\bm{W_{q}}$, $\bm{U_{q}}$, $\bm{W_{r}}$, $\bm{U_{r}}$, $\bm{W_{\hat{u}}}$, $\bm{U_{\hat{u}}}$ are the weight matrices to learn,
$\bm{q}$ is the weighting vector between $\bm{s}$ and $\bm{\hat{u}}$,
and $\bm{\hat{u}}$ denotes the information used to expand $\bm{s}$.
The output $\bm{\hat{s}}$ replaces the original vector $\bm{s}$ to represent the source method $\mathcal{S}$ in \smnn.

\subsection{Fault Localization Network}
Given a bug report $\mathcal{R}$, \flnn combines the explicit relevance score $e$
obtained from \msf and the three \bff (\textit{rcfs}, \textit{bffs} and \textit{bfrs})
to predict the probability that method $\mathcal{S}$ causes the
faults described in bug report $\mathcal{R}$.
Specifically, these four inputs are first concatenated as a feature vector $\bm{z}$. For better feature interaction, $\bm{z}$ is projected into a continuous vector by MLPs.

\begin{equation}
	\hat{y_{sr}} = \mbox{Softmax}(\bm{W_{2}}\sigma (\bm{W_{1}} \bm{z})+b_{1}) + b_{2}) ,
\end{equation}
where $\bm{W_{i}}$ and $b_{i}$ are the weight of the $i^{th}$ MLP layer,
$\sigma$ is an activation function,
and $\hat{y_{sr}}$ represents a prediction probability.

\subsection{Model Training}
Now we present how to train the \model to locate the faulty methods for a given bug report.
The high-level goal is:
given a bug report $\mathcal{R}$ and an arbitrary method $\mathcal{S}$
with its \bff $\mathcal{F}$,
we want \model to predict a high relevance score if $\mathcal{S}$ is a faulty method for $\mathcal{R}$,
and a low relevance score otherwise.

We construct each training instance as a triple
$\langle$$\mathcal{S}$, $\mathcal{F}$, $\mathcal{R}$$\rangle$
and reduce the task to a binary classification problem.
The binary cross-entropy loss function is defined as follows:
\begin{equation}
	\mathcal{L} = -\frac{1}{n}\sum_{(s,r)\in \mathbb{D}}(y_{sr}\ log \ y_{sr}- (1-y_{sr})log(1-y_{sr})) ,
\end{equation}
where $\mathbb{D}$ is the training dataset and
$y_{sr} \in \left\{ 0, 1 \right\}$ represents whether the source method $\mathcal{S}$
is fixed by the bug report $\mathcal{R}$.
Intuitively, the ranking loss encourages the relevance between a bug report and its fixed methods to go up,
and the relevance between a bug report and methods not fixed by it to go down.

\section{Experiment}

We conduct a series of experiments on five open source Java projects,
with the aims of answering the following questions:

\begin{itemize}
	\item \textbf{RQ1:} How effective is \model in faulty methods localization task, compared with state-of-the-art techniques?
	\item \textbf{RQ2:} How do different components contribute to the fault localization performance of \model ?
	\item \textbf{RQ3:} How does \model perform in cross-project fault localization?
\end{itemize}

\subsection{Experimental Setup}

\noindent\textbf{Baselines.}
We compare \model with the following state-of-the-art techniques.

\begin{enumerate}

\item \fine \cite{FineLocator} :
It is a state-of-the-art method level fault localization approach proposed recently.
It augments short methods by their neighboring methods according to
three query expansion scores.


\item \emph{\blia}\cite{BLIA} :
It 
integrates stack trace feature, collaborative filtering score,
and commit history feature with structural VSM and achieves better results than previous IR-based approaches like BugLocator \cite{buglocator}.

\item \emph{\bliz}\cite{Blizzard} :
It locates faults by employing context-aware query reformulation and information retrieval.
It 
applies appropriate reformulation techniques to refine the bug report in poor quality.
Then, the improved bug report is used for the fault localization with Lucene.
\end{enumerate}

\smallskip\noindent\textbf{Dataset.} We reuse the benchmark dataset of Ye \etal~\cite{fse2014}. Pendlebury \etal~\cite{pendlebury2019tesseract} criticize that it introduces bias if a model is trained from future data and tested on past data. To resolve the problem, we sort the bug reports chronologically, and divide the bug reports of each project into 10 folds with equal sizes. We train the model on $fold_{i},fold_{i+1},fold_{i+2}$
and test it on $fold_{i+3}$ to ensure that we have sufficient data for training. In addition, when we collect the buggy commit of a bug report, we select its before-fix commit rather than the commit when a report is issued as used by \cite{DNNLoc}, since a found bug shall appear at a before-fix commit.

What's more, we notice that it is impossible to use all the negative samples for training as negative samples are much more than positive ones.
For example, the latest revision of Birt has more than 100,000 methods (see Table \ref{dataset} for detail), and each bug report can have the same amount of  negative samples. To reduce the bias, we randomly select 300 negative samples for each bug report when we train the model, but we use all the samples when we test our model.


\smallskip\noindent\textbf{Metrics.}
We use three metrics to evaluate the performance of all approaches, namely, \textit{top-ranked accuracy (Top@k)},
\textit{mean average precision (MAP)}, and \textit{mean reciprocal rank (MRR)}, which have been widely used in fault localization and information retrieval~\cite{fse2014, DNNLoc,FineLocator,BLIA}.

\textit{Top@k} indicates the percentage of bug reports for which at least one of its fixed method in the \textit{Top-k} ranked results.
It is calculated as follows:
\begin{equation}
	Top@k = \frac{1}{|R|}\sum_{r=1}^{|R|}\delta(FRank_{r} \leq k) ,
\end{equation}
where $|R|$ is the number of bug reports, $FRank_{r}$ is the rank of the first hit faulty method $r$ in the result list \cite{raghothaman2016swim},
and $\delta$ is a function which returns 1 if the input is true and 0 otherwise.

\textit{MAP} measures the ability of fault localization approach to consider the bug report with multiple fixed methods. It is calculated as follows:
	\begin{equation}
	\begin{split}
		avgPre &= \frac{1}{|M|}\sum_{m=1}^{|M|}\frac{m}{Rank_{m}} ,\\
		MAP &= \sum_{r=1}^{|R|}\frac{avgPre_{r}}{|R|} ,
	\end{split}
	\end{equation}
where $|M|$ presents the number of fixed methods for a bug report,
and $Rank_{m}$ is defined as rank of the $m^{th}$ faulty method.

\textit{MRR} concentrates on the rank of the first fixed method in the ranked list, 
and it is calculated as follows:
\begin{equation}
	MRR = \sum_{r=1}^{|R|} \frac{1}{FRank_{r}}.
\end{equation}


\subsection{RQ1: The Improvements}
\begin{table*}[ht]
	\caption{The Top@k values of \model and the baselines}\vspace{-1mm}
	\begin{center}
	\footnotesize	
	\begin{tabular}{m{0.7cm}<{\centering}cccccccccccccc}
	\toprule
	\multirow{2}*{\textbf{Project}}
	& \multirow{2}*{\textbf{Approach}}
	& \multicolumn{2}{c}{Top@1}
	& \multicolumn{2}{c}{Top@5}
	& \multicolumn{2}{c}{Top@10}
	& \multicolumn{2}{c}{MAP}
	& \multicolumn{2}{c}{MRR} \\
	\cline{3-12}
	& &w(\%) &w/o(\%)  &w(\%) &w/o(\%)  &w(\%)&w/o(\%)  &w &w/o  &w &w/o  \\
	\midrule[1.1pt]
	\multirow{4}*{Tomcat}
	&\blia   &0 &0 &0.165 &0 &0.165 &0 &0.0004 &0.0001 &0.0004 &0.0001  \\
	&\bliz   &0 &0 &0.228 &0 &0.457 &0.248 &0.0032 &0.0003  &0.0031 &0.0003    \\
	&\fine   &0.667 &0.324 &2.667 &1.831 &4.833 &4.167 &0.0204 &0.0192 &0.0261 &0.0239  \\
	&\model  &\textbf{1.667} &\textbf{1.167} &\textbf{5.333} &\textbf{5.132} &\textbf{9.367} &\textbf{8.501} &\textbf{0.0378} &\textbf{0.0356} &\textbf{0.0492} &\textbf{0.0462}   \\
	\midrule
	\multirow{4}*{AspectJ}
	&\blia  &1.773 &3.200 &6.028 &5.600 &7.447 &7.200 &0.0343 &0.0419 &0.0391 &0.0480  \\
	&\bliz  &0.781 &0.800 &3.385 &2.400 &4.687 &3.200 &0.0173 &0.0154 &0.0183 &0.0154  \\
	&\fine  &1.563 &1.035 &7.188 &6.863 &9.688 &8.753 &0.0347 &0.0402 &0.0503 &0.0471 \\
	&\model &\textbf{5.937} &\textbf{4.438} &\textbf{13.563} &\textbf{11.751} &\textbf{16.123} &\textbf{15.753} &\textbf{0.0742} &\textbf{0.0686} &\textbf{0.0921} &\textbf{0.0847}  \\
	\midrule
	\multirow{4}*{SWT}
	&\blia   &3.409 &2.465 &5.681 &5.343 &6.818 &6.849 &0.0455 &0.0383 &0.0457 &0.0399  \\
	&\bliz &5.825 &2.336 &15.426 &7.476 &19.633 &9.345 &0.1015 &0.0463 &0.1063 &0.0497  \\
	&\fine &4.632 &3.082 &7.634 &5.733 &9.232 &6.032 &0.0502&0.0456 &0.0533 &0.0483
	\\
	&\model &\textbf{9.698} &\textbf{7.113} &11.422 &\textbf{8.763} &14.116 &\textbf{11.321} &0.0987 &\textbf{0.0765} &\textbf{0.1114} &\textbf{0.0838} \\
	\midrule
	\multirow{4}*{JDT}
	&\blia   &2.319 &1.076 &6.714 &2.929 &9.155 &5.020 &0.0422 &0.0214 &0.0461 &0.0236   \\
	&\bliz &0 &0 &0.095 &0.060 &0.151 &0.179 &0.0005 &0.0004 &0.0005 &0.0004  \\
	&\fine &0.917 &0.341 &3.511 &2.843 &6.710 &4.672 &0.0238 &0.0183 &0.0335 &0.0205  \\
	&\model &\textbf{4.613} &\textbf{3.672} &\textbf{7.951} &\textbf{6.932} &\textbf{12.920} &\textbf{12.513} &\textbf{0.0491} &\textbf{0.0391} &\textbf{0.0702} &\textbf{0.0627}  \\
	\midrule
	\multirow{4}*{Birt}
	&\blia   &1.576 &0.883 &5.086 &3.092 &6.698 &4.290 &0.0299 &0.0189 &0.0331 &0.0205   \\
	&\bliz &0 &0 &0.028 &0.063 &0.028 &0.063 &0.0002 &0.0003 &0.0002 &0.0003  \\
	&\fine &1.632 &0.023 &3.447 &1.313 &5.324 &2.221 &0.0266 &0.0083 &0.0293 &0.0095 \\
	&\model &\textbf{3.755} &\textbf{3.322} &\textbf{6.321} &\textbf{7.631} &\textbf{10.443} &\textbf{9.572} &\textbf{0.0434} &\textbf{0.0423} &\textbf{0.0592} &\textbf{0.0561}  \\
	\midrule[1.1pt]
	\multirow{4}*{\textbf{Average}}
	&\blia  &1.815 &1.525 &4.735 &3.393 &6.057 &4.672 &0.0305 &0.0241 &0.0329 &0.0264 \\
	&\bliz &1.321 &0.627 &3.832 &2.000 &4.991 &2.596  &0.0249 &0.0125 &0.0257 &0.0132  \\
	&\fine &1.882 &0.961 &4.889 &3.717 &7.157 &5.169 &0.0311 &0.0263 &0.0385 &0.0299  \\
	&\model &\textbf{5.134} &\textbf{3.942} &\textbf{8.918} &\textbf{8.042} &\textbf{12.594} &\textbf{11.532}
	&\textbf{0.0606} &\textbf{0.0524} &\textbf{0.0764} &\textbf{0.0667}  \\
	\bottomrule
	\end{tabular}		
	\label{comparison}
	\vspace{1mm}
	\begin{tablenotes}
		\item[a] * w stands for using all bug report for evaluating; w/o stands for using only \textit{not localized} bug report for evaluating.
	\end{tablenotes}
	\end{center}
	\vspace{-2mm}
\end{table*}

To answer this RQ, we compare the effectiveness of \model with state-of-the-art fault localization techniques on the dataset with all bug reports and only \textit{not localized} bug reports.
Table \ref{comparison} shows the overall performance of \model and baselines,
measured in terms of Top@1/5/10, MAP and MRR. 

We have observed that \model achieves more promising overall fault localization results than other techniques.
For example, the average Top@1 value of \model is 5.134 much higher than \blia's 1.815, \bliz's 1.321, and \fine's 1.882.
Also, the MAP and MRR values of \model are the best among all studied techniques. 
One reason for this is that \model considers both the implicit and explicit relevance between bug reports and source methods, 
providing more effective fault-diagnosis information analysis.

We have also observed that the performance of all techniques is decreased after filtering out localized bug reports,
which could lead a fault localization technique to be not as effective in practice as in experiments.
It is worth noting that although the performance of our \model is also decreased, it only drops slightly compared to other techniques.
For example, the MRR value of \model drops 12.696\%,
compared to \blia's 19.757\%, \bliz's 48.638\%, and \fine's 22.338\%.
One potential reason is that when localized bug reports are filtered out,
it is difficult for \blia and \bliz to utilize the keywords (\ie method name)
existed both in bug reports and their fixed methods to compute a high textual similarity.
\bliz which relies extremely on those keywords drops the most. 
For example, in SWT, nearly 60\% bug reports (see Table~\ref{dataset}) are localized.
Therefore, it is easy to locate their faulty methods by matching their method names with \bliz. However, a new bug report may not contain so many method names, and its effectiveness is thus reduced. 

\blia integrates other sources information besides textual similarity, 
and therefore reduces the impacts brought by filtering keywords compared to \bliz.
Both \model and \fine rely on embedding technique, and thus elegantly avoid this problem.
Moreover, compared to \fine, which treats methods as plain texts,
\model considers \msf and incorporates the relevant information by focusing on 
the representation of bug reports, and thus performs better. Therefore, we have the first finding:
\begin{framed}
\noindent\emph{Finding 1}:
Comparing with state-of-the-art approaches, \model improves MRR values by 3.8-5.1\% (3.7-5.4\%), under the settings with (without) localized bug reports.

\end{framed}

It is worth noting that the performance of all models in Tomcat is not promising. 
With manual analysis on bug reports of the Tomcat project, 
we found that, different from other projects, there are a certain number of bug reports 
that are raised by the developers themselves to record new features. 
They are actually not ``bug reports'', 
and thus could not be localized by fault localization techniques. 
We consider it as future works to identify these bug reports automatically, 
and analyze the influence of these bug reports during model training and evaluation.

To investigate our improvements, we further use the Mann-Whitney U Test\cite{mann1947test} to compare our Top@k values with those of the prior approaches. The results show that \model is significantly better than all compared approaches in terms of Top@k at significance level of 0.05. 
In particular, the p-values varies from 0.00110 to 0.00367 when the dataset contains localized bug reports
and p value varies from 0.00013 to 0.00162 when the dataset does not contain such reports.

\subsection{RQ2: Our Internal Techniques }
\begin{table*}[htp]
	\caption{The impacts of the internal techniques of \model (the SWT project)}\vspace{-2mm}
	\begin{center}
		\scalebox{1}{
		\begin{tabular}{ccccccccccc}
			\toprule
			\textbf{Model} &Top@1(\%) &$\triangle$ &Top@5(\%) &$\triangle$ &Top@10(\%) &$\triangle$  &MAP  &$\triangle$ &MRR &$\triangle$ \\
			\midrule
			\model                &9.698 &-      &11.422 &- &14.116 &- &0.0984 &- &0.1114 &- \\

			\model-\textit{rcfs}  &5.567 &-4.131 &9.897  &-1.525 &10.619 &-3.497 &0.0660 &-0.0324 &0.0738 &-0.0376  \\
	
			\model-\textit{bffs}  &7.113 &-2.585 &8.763  &-2.659 &9.381  &-4.735  &0.0674 &-0.0310 &0.0816 &-0.0298 \\
	
			\model-\textit{bfrs}  &6.907 &-2.791 &7.732  &-3.690 &8.557  &-5.559  &0.0646 &-0.0338 &0.0778 &-0.0336 \\
			
			\model-SMNN           &8.632 &-1.066 &10.413 &-1.009 &12.124 &-1.992  &0.0893 &-0.0091 &0.1052 &-0.0062 \\

			\model-graph         &6.846 &-2.852 &8.631 &-2.791 &10.577 &-3.593 &0.0713 &-0.0271 &0.0769 &-0.0345 \\
			\bottomrule
		\end{tabular}
		}
		\label{ablation}
	\end{center}
	\vspace{-5mm}
\end{table*}

To get a better insight into \model, 
an in-depth ablation study is conducted on SWT project. 
The main goal is to validate the effectiveness of the critical features or components in our architecture including
revised collaborative filtering score (\textit{rcfs}), bug-fixing frequency score (\textit{bffs}), bug-fixing recency score (\textit{bfrs}), 
method expansion network (\menn), and code revision graph.
When without the code revision graph, we use only the latest code revision of SWT to compute the features of methods. 

As Table \ref{ablation} shows, we can find that: 

1) All the investigated components in \model are helpful in locating faulty methods, and
among them \textit{rcfs} contributes mostly on Top@1.
The benefit of \textit{rcfs} lies in two parts:
a) it gets more accurate explicit relevance between bug reports and source methods by 
revising original collaborative filtering score;
b) it improves the localization of short methods significantly by exploiting the relations in \graph. 

2) Both \textit{bffs} and \textit{bfrs} are vital for the final performance. 
This indicates the pattern that the frequently fixed methods and recently fixed methods
in the past are the error-prone ones.

3) \menn also benefits for the performance of \model to some extent, 
by alleviating the representation sparseness problem. 
However, it seems that it works not as well as other components.
The possible reason is that relations between methods have already been exploited in \textit{rcfs} more explicitly.

4) Code revision graph plays an important part in locating faulty methods. Using only the latest code revision may result in the loss of method historical information, 
which is caused by code changes (\ie change method signature, deleting method) across code revisions. Thus the calculated \bff are not complete and accurate to train a model. Then we have the second finding:
\begin{framed}
	\noindent\emph{Finding 2}:
 Code revision graphs are useful to resolve the single revision problem, and \textit{rcfs} and MENN are useful to handle the sparseness representation problem.
\end{framed}

\subsection{RQ3: Cross-project Localization}

The localization in previous RQs are within-project.
To further investigate the effectiveness of \model, 
we extend the fault localization to cross-project scenarios, which will benefit a startup project without enough training data to locate faulty code precisely.
Since the experiment is time-sensitive, instead of using k-fold cross validation like prior work\cite{deepFL},
the data of one project is used as the training data while the
data from another project is treated as the testing data.

\begin{table}[t]
	\caption{The results of learning from other projects}\vspace{-1mm}
	\begin{center}
		\scalebox{1}{
		\begin{tabular}{p{1cm}<{\centering} p{1cm}<{\centering} p{1cm}<{\centering} p{1.2cm}<{\centering} p{1cm}<{\centering} p{1cm}<{\centering}}
			\toprule
			\textbf{Task} &Top@1(\%) &Top@5(\%) &Top@10(\%) & MAP(\%) &MRR(\%)  \\
			\midrule

			B$\rightarrow$A &4.063 &6.563 &11.875 &0.0437 &0.0600 \\

			J$\rightarrow$B  &2.926 &5.535 & 8.067 &0.0353 &0.0502 \\

			S$\rightarrow$J  &3.676  &7.021  &9.176  &0.0396  &0.0556 \\

			T$\rightarrow$S &5.872 & 8.313 &11.072 &0.0631 &0.0783  \\

			A$\rightarrow$T &1.332 &3.434 &7.591 &0.0215 &0.0306 \\
			\midrule
			\textbf{Average} &3.574 &6.173 &9.556 &0.0406 &0.0549 \\ 

			\bottomrule
		\end{tabular}}
		\label{cross-project}
	\end{center}
	\vspace{-4mm}
	\end{table}
Table \ref{cross-project} presents the experimental results of \model for
the cross-project localization.
The results show that the average MAP and MRR values of cross-project localization are 0.0406 and 0.0549, respectively,
slightly worse than within-project localization.
This is as expected: in within-project localization, the training data and test data tend to have similar data distribution 
since they come from the same project,
while in cross-project localization, the data distributions are rather different.
We also observe that our \model in cross-project scenario
still outperforms the compared techniques in within-project scenario.
This indicates that our model can effectively utilize the data from a project to train a model 
for another project which has insufficient labelled bug reports.
So we have the final finding:

\begin{framed}
	\noindent\emph{Finding 3}:
	In the cross-project setting, the effectiveness of \model is only slightly worse than that of the within-project setting.
\end{framed}

\subsection{Threats to Validity}
For internal validity, 
the main threat is the potential mistake in our technique implementation and features collection. 
To reduce this threat, we implement our approach by utilizing state-of-the-art frameworks and tools,
such as Pytorch\cite{paszke2019pytorch}, Spoon\cite{pawlak2016spoon}, and Neo4j\cite{neo4jurl}.
For external validity, the main threat is the selection of the studied projects.
To mitigate this threat, we evaluate all approaches on five open-source projects, 
which are used in the evaluations of the prior techniques\cite{fse2014, DNNLoc}.
These projects are not only developed by different developers with different programming idioms
but also different in size, thus can reflect the real-world situations.
In the future, we will extend the scale and scope of studied projects. 
For construct validity, the main threat is that the metrics used may not fully reflect real-world situations. 
To reduce this threat, we use Top@1/5/10, MAP, and MRR metrics, which have been widely used in the previous works\cite{buglocator, BLIA, DNNLoc, FineLocator}.
Our measures consider only the top 10 recommendations, while a lower recommendation can still be a real faulty method.
However, we believe that the setting is reasonable. 
As developers often do not inspect more than top 10 results \cite{practitioner}, it does not make much difference if a method appears at rank 11 or later.

\section{Related Work}

\noindent \textbf{IR-based Fault Localization.} 
One line of works take bug reports as their inputs. 
Most of them adopt various machine learning techniques to detect the similarity between bug reports and source code \cite{kim2013should,lukins2010bug,rao2011retrieval,wang2014compositional}.
Kim \etal \cite{kim2013should} transformed the bug report into a feature vector and
used naive Bayes to match it with source files.
Lukins \etal \cite{lukins2010bug} showed the performance of LDA-based fault localization model is 
not affected by the size of subject software system.
These works mostly treated code snippets as the natural language by ignoring the structural information of the program.
More recent studies ~\cite{saha2013improving} extracted structured information (\ie class name, method name, and comment) from source code to improve fault localization.
Wang \etal \cite{wang2015evaluating} showed these keyword-based techniques rely heavily on the code names that appear in both bug reports and source code, which may lead to a marked decline in performance when localized bug reports are filtered out\cite{bias2014}. 
Zhang \etal ~\cite{zhang2019cnn} introduced deep learning networks and Huo \etal \cite{8736995} adopted deep transfer learning to fault localization. Besides matching textual similarity between bug reports and source code, other data source (\eg stack traces\cite{wong2014boosting}, version histories \cite{sisman2012incorporating}) are also investigated.
Akbar \etal \cite{akbar2019scor} proposed a code retrieval framework to locate faults.

Differing from existing techniques,
\model measures both implicit relevance and explicit relevance between bug reports and source methods.
To calculate implicit relevance, it does not rely on information retrieval techniques,
but uses deep learning techniques to cleverly integrate program structure to represent source methods and
match them with bug reports in the same vector space. Thus, it could better understand the semantics of methods and bug reports.
It calculates \bff on all code revisions rather than the latest code revision used in previous techniques.
Therefore, the calculated explicit relevance is more accurate and effective.

\smallskip\noindent \textbf{Spectra-based Fault Localization.} The approaches
in another line calculate the suspicious scores of program elements using the program's execution information. They require an executable system with many high-quality test cases, which are often unavailable when a bug was reported~\cite{ifixR}. Failed test cases and program behavior traces are used for fault localization in \cite{korel1988stad, taha1989approach, agrawal1991execution}.
Renieres \etal \cite{renieres2003fault} concerned on the source code only executed by the failed test or by all the successful tests, which are considered more suspicious.
Liblit \etal \cite{liblit2005scalable} proposed a statistical debugging technique to isolate bugs in programs with instrumented predicates.
Zimmermann \etal \cite{zimmermann2002visualizing} presented a program state-based technique that
contrasted program states between executions of a successful test and a failed test using memory graphs.
Lo \etal \cite{deepFL} applied deep learning techniques to analyze the program execution information of passed and failed test cases.
However, test cases are always not available when bugs are reported,
which makes those \textit{spectra-based fault localization} approaches not suitable for time-critical debugging tasks \cite{ifixR}.

\smallskip\noindent \textbf{Source Code Representation in Deep Learning.} 
Recently, researchers have investigated a series of techniques to represent source code by using deep learning techniques to improve the performance of various code intelligence tasks
\cite{astnn,gu2018deep,cao2020bugpecker,li2020learning,huo2016learning,7582748}.
Zhang \etal \cite{astnn} utilized split abstract syntax tree (AST) to capture lexical and syntactical knowledge of code fragment and achieved the best performance on two common program comprehension tasks: source code classification and code clone detection.
White \etal \cite{7582748} proposed a recursive
neural network (RvNN) to link patterns mined at the lexical level and a recurrent neural network (RtNN) to link patterns mined at the syntactic level for code clone detection.
Gu \etal \cite{gu2018deep} jointly embedded source code and natural language queries into a unified space by using RNN to measure their semantic similarity for code search task.
Instead of fusing multiple structured information directly, 
our proposed \model takes the bug reports as a reference to integrate token sequence,
API invocation sequence and comment to represent the source methods, by using a soft attention mechanism.
Moreover, we construct \graph from code, commits and past bug reports, 
and reveal the latent relationships among methods to augment short methods before method embedding, 
and thus alleviate the representation sparseness problem.

\section{Conclusion}
In this paper, we propose a mixed deep neural network named \model for method level fault localization task.
\model learns a unified vector representation of both source methods and natural language bug reports and matches them by vector similarities.
Instead of only using the latest code revision, \model constructs \graph from multiple code revisions and past fixes. Based on such graphs, \model further expands short methods and calculates \bff to improve the effectiveness of locating faults at method level.
As a proof-of-concept application, we implement a faulty method locating tool based on \model. We compare \model with three state-of-the-art approaches on a widely used dataset. Our experimental results show that \model is effective and outperforms the compared approaches.

In the future, we will investigate to capture program structure 
(\eg AST, control flow or data flow)  to better represent the source methods. To improve the evaluation, we plan to apply \model to locate real open bug reports in both open-source projects and industry projects.

\section{Acknowledgment}
This research is supported by National Natural Science Foundation of China (Grant No. 62032004) and National Key Research and Development Program of China (Grant No. 2018YFB1003903).

\balance
\bibliographystyle{IEEEtran}
\bibliography{DFCNN}

\end{document}